\documentclass[a4paper,10pt]{article}

\usepackage{amsmath}
\usepackage{amsfonts}
\usepackage{amssymb}
\usepackage{color}

\newcommand{\be}{\begin{equation}}
\newcommand{\ee}{\end{equation}}
\newcommand{\bea}{\begin{eqnarray}}
\newcommand{\eea}{\end{eqnarray}}
\newcommand{\nn}{\nonumber}

\newcommand{\one}{\mbox{\bf 1}}
\newcommand{\eps}{\epsilon}
\newcommand{\al}{\alpha}
\newcommand{\la}{\lambda}
\newcommand{\Ga}{\Gamma}
\newcommand{\bD}{\bar{\Delta}}
\def\Tr{{\mbox{Tr}}}

\def\hla{\hat{\lambda}}

\newcommand{\sect}[1]{\setcounter{equation}{0}\section{#1}}

\begin{document}

\title{
Finite size scaling of meson propagators with\\ isospin chemical potential
}

\author{~\\{\sc G.~Akemann}$^{1}$, {\sc F.~Basile}$^{1,2}$ and 
{\sc L. Lellouch}$^{3}$ 
\\~\\
$^1$Department of Mathematical Sciences \& BURSt Research Centre\\
Brunel University West London,
Uxbridge UB8 3PH, United Kingdom
\\~\\
$^2$Dipartimento di Fisica dell'Universit\`a di Pisa \& INFN\\
Largo B. Pontecorvo 3, 56127 Pisa, Italy
\\~\\
$^3$Centre de Physique Th\'eorique, CNRS Luminy,\\
Case 907,
F-13288 Marseille Cedex 9, France\footnote{CPT is 
``UMR 6207 du CNRS et des universit\'es d'Aix-Marseille I, d'Aix-Marseille II
  et du Sud Toulon-Var, affili\'ee \`a la FRUMAM''}}

\maketitle 

\begin{abstract}

  We determine the volume and mass dependence of scalar and
  pseudoscalar two-point functions in $N_f$-flavour QCD, in the
  presence of an isospin chemical potential and at fixed gauge-field
  topology. We obtain these results at second order in the
  $\eps$-expansion of Chiral Perturbation Theory and evaluate all
  relevant zero-mode group integrals analytically.  The virtue of
  working with a non-vanishing chemical potential is that it provides
  the correlation functions with a dependence on both the chiral
  condensate, $\Sigma$, and the pion decay constant, $F$, already at
  leading order. Our results may therefore be useful for improving the
  determination of these constants from lattice QCD calculations.  As
  a side product, we rectify an earlier calculation of the
  $\mathcal{O}(\eps^2)$ finite-volume correction to the decay constant
  appearing in the partition function. We also compute a generalised
  partition function which is useful for evaluating $U(N_f)$ group
  integrals.

\end{abstract}
\vfill 
 
\begin{flushright} 
CPT-P004-2008
\end{flushright} 
\thispagestyle{empty} 

\newpage

\sect{Introduction}

With the possibility of performing unquenched lattice gauge theory
calculations at increasingly small quark masses, new opportunities for
understanding the low-energy properties of QCD are emerging. The
theoretical framework for such analyses is provided by Chiral
Perturbation Theory (ChPT), which describes the long wavelength
excitations of QCD above the vacuum, in terms of physical parameters
known as low-energy constants (LECs). While there exist a number of
ways to approach the chiral limit, we will consider here the
$\eps$-regime of \cite{GL87a}. This regime is well suited to numerical
lattice calculations, since it is reached by lowering the quark masses
in such a way that the Compton wavelength of the corresponding
pseudo-Goldstone bosons is kept much larger than the typical linear extent
of the finite (simulated) four-volume $V$. Low energy constants can be
determined in this regime, by comparing lattice calculations of a
given observable with analytical expressions for this observable
obtained in ChPT.  In the present work we will focus on ChPT
calculations which allow the determination of the two leading LECs,
the chiral condensate $\Sigma$ and the pion decay constant $F$, and we
refer to \cite{Necco} for a recent review of lattice results for these
as well as higher order LECs.

Two-point correlation functions, of either scalar and pseudoscalar
densities, or of vector and axial vector currents, are relatively
straightfoward quantities to compute with lattice QCD methods.  They
are also calculable in the $\eps$-regime of ChPT, and have been
studied up to and including second order corrections, in the quenched,
partially quenched or full theory, with or without fixed gauge-field
topology
\cite{Hansen90,DDHJ,Damgaard:2002qe,Damgaard:2007ep,BernardoniHernandez}.
Comparisons of these analytical predictions with lattice results have
been reported in \cite{Lattice}.  At second order, these correlation
functions couple to both LECs, $\Sigma$ and $F$. However, in scalar
and pseudoscalar two-point functions, $F$ only appears at sub-leading
${\cal O}(\eps^2)\sim 1/\sqrt{V}$. In vector and axial correlation
functions, it is present at leading non-vanishing order, but these
functions are $\mathcal{O}(\eps^4)$-suppressed compared to the scalar
and pseudoscalar correlators. All of these suppression factors are
clearly a handicap for determining $F$ from lattice calculations of
the above two-point functions.

For that reason, we propose to consider two-point scalar and
pseudoscalar correlation functions in the presence of a non-zero
imaginary isospin chemical potential $\mu$, leaving vector and
axial-vector correlation functions for future work. The interest of
adding a chemical potential is that it couples the decay constant,
$F$, to the zero modes in the leading order Lagrangian, through the
combination $\mu^2VF^2$ \cite{Toublan}. This implies that the scalar
and pseudoscalar correlation functions now depend on $F$, already at
leading order. In turn, the increased sensitivity to $F$ should
facilitate the extraction of this LEC from lattice calculations. Since
all of our results depend only on $\mu^2$, they hold true for both
real and imaginary isospin chemical potentials for any number of
flavours $N_f$. For real $\mu$, only the even flavour case has a real
action.

The proposal to use a chemical potential to facilitate the
determination of $F$ was originally made in the context of spectral
correlators \cite{Imiso,AW}. In \cite{Imiso}, the spectral two-point
density was singled out as particularly sensitive to $F$. More
recently, all spectral correlators \cite{ADOS} and individual
eigenvalue correlations \cite{AD08} with imaginary isospin chemical 
potential were computed in chiral Random
Two-Matrix Theory, which is equivalent to leading order ChPT in the epsilon
regime \cite{BA}. This setup includes the partially quenched theory,
where $\mu$-dependent valence quarks propagate in a background of
$\mu$-independent sea quarks, allowing for the use of standard,
zero-density lattice configurations. In fact, a lattice calculation of
$F$ has already been performed in this framework \cite{Tom}.

The remainder of the article is organised as follows. In
Sec.~\ref{Zeps}, we derive the ${\cal O}(\eps^2)$ improved partition
function and two-point meson correlation functions in terms of
expectation values over the pion zero-mode.  In particular, we correct
an earlier calculation \cite{DGF} of the one-loop finite-volume corrections
to the partition function.  In Sec.~\ref{Results} we present our
results for both flavoured and unflavoured scalar and pseudoscalar
two-point correlation functions in $N_f$-flavour QCD, given in terms
of the finite-volume propagator, the mass dependent condensate and
other generating functions. These general expressions are spelled out
most explicitly in the case of two flavours.  Our conclusions are
given in Sec.~\ref{conc}.  Two appendices, \ref{B} and \ref{C},
provide technical details for group integral identities, including a
new result for a generalised partition function.

\sect{The epsilon expansion in ChPT with a chemical potential}
\label{Zeps}

Since imaginary chemical potentials couple to fields like the time
component of a constant vector current, they can be accounted for in
ChPT through the covariant derivative on the pion fields, $U(x)\in SU(N_f)$:
\be
\nabla_\rho U(x)\equiv \partial_\rho U(x) - i\, \delta_{\rho,0}[C,U(x)]
\ ,\ \ 
\nabla_\rho U(x)^\dag
\ =\ \partial_\rho U(x)^\dag - i\, \delta_{\rho,0}[C,U(x)^\dag] \ ,
\label{nabladef}
\ee
where C is a matrix proportional to the chemical potential $\mu$,
whose form depends on the charge $\mu$ couples to. Moreover, it was
shown in \cite{GL87a}, for calculations up to and including second
order in the $\eps$-expansion, that the usual leading order chiral
Lagrangian is sufficient. Thus, we consider here the Lagrangian:
\be 
{\cal L}_2= \frac{F^2}{4}\Tr\left[\nabla_\rho
    U(x)^\dag\nabla_\rho U(x)\right] -\ \frac{\Sigma}{2}\Tr\left[
    \mathcal{M}^\dag U(x) +U(x)^\dag \mathcal{M}\right] \ ,
\label{L2}
\ee
where $\mathcal{M}$ is the mass matrix. We will take $\mathcal{M}=$
diag$(m_1,\ldots,m_{N_f})$ and throughout this section the external
vector current, $C$, will be kept general.  In Sec.~\ref{Results} we
will specify this current to be the isospin charge matrix, $C=$
diag$(\mu\one_{N_1},-\mu\one_{N_2})\equiv\mu\Ga$.  The simplest and
interesting case of $N_f=N_1+N_2=1+1$ flavours of equal mass
$m_{1,2}=m$ will be displayed in great detail in Sec.~\ref{Results}.
As our final results will depend only on $\mu^2$, we can also rotate 
back to real chemical potential by $\mu^2\to-\mu^2$ at the end of our
calculations, for any number of flavours  $N_f=N_1+N_2$. 

We use the standard parametrisation of the Goldstone fields for
the $\eps$-expansion:
\be
U(x)\equiv U_0\exp\left[i\frac{\sqrt{2}}{F}\xi(x)\right]\ .
\label{Udef}
\ee 
These fields live on the Goldstone group manifold, $U(x)\in\ SU(N_f)$,
whereas the propagating Hermitean fields $\xi(x)=\xi(x)^\dag$ are
members of the corresponding Lie algebra, e.g. for two flavours,
$N_f=2$, they are given by $\xi(x)=\frac12\sigma_a \xi_a(x)$, in terms of
the Pauli matrices.  In particular, we have split off the zero
momentum mode, $U_0\in\ SU(N_f)$, explicitly, and will treat its
effects exactly.  The power counting in the epsilon
expansion in the presence of a chemical potential is given
by~\cite{GL87a}:
\be
V\sim\epsilon^{-4},\quad \mathcal{M}\sim \epsilon^4,\quad\mu\sim \epsilon^2,
\quad\partial_\rho\sim \epsilon,\quad\xi(x)\sim \epsilon
\ .
\label{eq:powercounting}
\ee
All other quantities, such as $\Sigma$ and $F$, are of $\mathrm{O}(\eps^0)$.

With this power counting, it is straightforward to expand the action,
$\mathcal{S}_2\equiv \int\,d^4x\,
\mathcal{L}_2$ to $\mathrm{O}(\eps^2)$. With the notation
$\mathcal{S}_2={\cal S}^{(0)}+{\cal S}^{(1)}+{\cal
  S}^{(2)}+\mathcal{O}(\eps^3)$, we obtain:
\begin{itemize}
\item{${\cal O}(1)$:} \ \ \ \ \ ${\cal S}^{(0)}= {\cal S}_{U_0}^{(0)}+ {\cal
  S}_{\partial^2}^{(0)}$ 
\bea
 {\cal S}_{U_0}^{(0)}(\Sigma,F)&=& -V \frac{F^2}{4}
 \Tr\Big[[C,U_0^\dag][C,U_0] \Big] 
- V\frac{\Sigma}{2}\Tr[ \mathcal{M}^\dag U_0+U_0^\dag\mathcal{M}]
\label{eq:S0U0}\\
{\cal S}_{\partial^2}^{(0)}&=&\frac12 \int d^4x\;
\Tr[\partial_\rho\xi(x)\partial_\rho\xi(x)]\label{eq:S0d2} 
\eea
\item{${\cal O}(\eps)$:}
\be
{\cal S}^{(1)} =
-\frac{i}{2} \int d^4x\; \Tr\Big[\partial_0\xi(x)[C+U_0^\dag CU_0,\xi(x)]\Big]
\label{eq:S1}
\ee
\item{${\cal O}(\eps^2)$:} \ \ \ \ \ ${\cal S}^{(2)}= {\cal
  S}_{\partial^2}^{(2)}+{\cal S}_{\xi,U_0}^{(2)}$ 
\bea
{\cal S}_{\partial^2}^{(2)} &=&
\frac{1}{12F^2}\int d^4x\; \Tr\Big[[\partial_\rho\xi(x),\xi(x)]
[\partial_\rho\xi(x),\xi(x)]\Big]
\label{eq:S2d2}\\
{\cal S}_{\xi,U_0}^{(2)}&=&
-
\frac{1}{\sqrt{2}F}\int d^4x\; \Tr\Big[ \xi(x)\partial_0\xi(x)\,\xi(x)
U_0^\dag[C,U_0]
\Big]
-\frac12 \int d^4x \;
\Tr\Big[U_0^\dag CU_0[\xi(x),[C,\xi(x)]]\Big]\nn\\
&&+\frac{\Sigma}{2F^2} \int d^4x\; 
\Tr[\mathcal{M}^\dag U_0\xi(x)^2+\xi(x)^2U_0^\dag\mathcal{M}]
\label{eq:S2xiU0}
\eea
\end{itemize}
where the superscripts specify the order in the $\eps$-expansion.
Note that all terms containing $C$ vanish for $C\propto\one_{N_f}$, i.e. pions
do not couple to baryon chemical potential.

\subsection{Partition function at ${\cal O}(\eps^2)$ and
  effective couplings}

The partition function for the $N_f$-flavour chiral theory is given by
\bea
{\cal Z} &\equiv& \int_{SU(N_f)} [d_H U(x)]
\,\exp[-{\cal S}]
\label{eq:Zdef}\\
&=& \int_{SU(N_f)} d_H U_0
\exp[-{\cal S}_{U_0}^{(0)}(\Sigma,F)]\,\mathcal{Z}_\xi(U_0)
\nn\ , 
\eea
where
\be
\mathcal{Z}_\xi(U_0)\equiv\int[d\xi(x)]\,J(\xi)\,\exp[{\cal
    S}_{U_0}^{(0)}(\Sigma,F)-{\cal S}] 
\label{eq:ZxiU0}
\ee
and where $\mathcal{S}$ is the chiral action at an, as of yet,
unspecified order. In Eq.~(\ref{eq:Zdef}) we integrate over the
Goldstone manifold with Haar measure $d_H U$ and in
Eq.~(\ref{eq:ZxiU0}), $J(\xi)=\{1-\frac{N_f}{3F^2V}\int
d^4x\Tr[\xi(x)^2]+\mathcal{O}(\eps^3)\}$ is the Jacobian corresponding
to the change of variables of Eq.~(\ref{Udef})~\cite{GL87a}.

In this theory, expectation
values are given by
\be 
\langle O \rangle_{U(x)}\ \equiv \ \frac{1}{\cal
    Z} \int_{SU(N_f)} [d_H U(x)]\ O\ \exp[-{\cal S}]\ ,
\label{Uxave}
\ee
As we are interested in evaluating observables up to ${\cal
  O}(\eps^2)$, we need the partition function at that order.  Since
${\cal S}_{U_0}^{(0)}(\Sigma,F)$ is $\mathcal{O}(1)$, we must compute
$Z_\xi(U_0)$ to $\mathcal{O}(\eps^2)$. Expanding $J(\xi)\exp[{\cal
  S}_{U_0}^{(0)}(\Sigma,F)-S]$ to second order and performing the
resulting Gaussian integrals with the propagator,
\be
\frac{1}{\int[d\xi(x)]e^{-{\cal S}_{\partial^2}^{(0)}}}
\int[d\xi(x)]e^{-{\cal S}_{\partial^2}^{(0)}}\
\xi(x)_{ij}\xi(y)_{kl}
\ =\ \Big(\delta_{il}\delta_{jk}-\frac{1}{N_f}\delta_{ij}\delta_{kl}\Big)
\bD(x-y)\ ,
\label{propdef}
\ee
we obtain
\bea \mathcal{Z}_\xi(U_0)&=&\mathcal{N}\left\{1-\frac{V\Sigma}{2F^2}
  \frac{(N_f^2-1)}{N_f}\bD(0)\,\Tr[\mathcal{M}^\dag
  U_0+U_0^\dag\mathcal{M})]\right.\nn\\
& & \qquad\left. -\frac{VN_f}{2}\Big(\bD(0)-\int d^4x
  [\partial_0\bD(x)]^2\Big)\,\Tr\Big[[C,U_0^\dag][C,U_0]
  \Big]\right\}+\mathcal{O}(\eps^3) \ ,
\label{eq:ZxiU0eps2}
\eea
where $\mathcal{N}$ is an overall normalisation factor which does not
contribute to the expectation values defined above. 
Now, re-exponentiating the corrections
in Eq.~(\ref{eq:ZxiU0eps2}), we obtain for the partition function:
\be \mathcal{Z} = \mathcal{N}\int_{SU(N_f)}d_H U_0\,\exp[-{\cal
  S}_{U_0}^{(0)}(\Sigma_{ef\!f},F_{ef\!f})] +\mathcal{O}(\eps^3)
\label{eq:Zeps2}
\ ,\ee
with $\Sigma$ and $F$ replaced in the argument of ${\cal
  S}_{U_0}^{(0)}$ by the 1-loop corrected couplings:
\bea
\Sigma_{ef\!f}&=& \Sigma
\left( 1-\frac{(N_f^2-1)}{N_fF^2}\bD(0)\right),
\label{Sigeff}\\
F_{ef\!f}&=& F
\left(1-\frac{N_f}{F^2}\Big(\bD(0)-\int d^4x [\partial_0\bD(x)]^2\Big) \right).
\label{Feff}
\eea

While the correction to the condensate $\Sigma$
Eq.~(\ref{Sigeff}) has been known a long time \cite{GL87a}, the
correction to $F$ was computed only very recently \cite{DGF}, apart
from the second term in Eq.~(\ref{Feff}) which is new and seems to
have been omitted in \cite{DGF}. This term arises from the
contribution proportional to $({\cal S}^{(1)})^2$ in the computation of
$\mathcal{Z}_\xi(U_0)$ to $\mathcal{O}(\eps^2)$.

In dimensional regularisation, the propagator $\bD(0)$ is finite and
is given by \cite{Hasenfratz}:
\be
\bD(0)\ =\ -\beta_1/\sqrt{V}\ .
\label{correctiondef}
\ee 
Moreover \cite{Hansen90},
\be
\int d^4x\, [\partial_{0}{\bD}(x)]^2
\ =\ -\frac1{2\sqrt{V}}\left[\beta_1-\frac{T^2}{\sqrt{V}}k_{00}\right]\ ,
\label{Z_eps2}
\ee
where $T$ is the time extent of the box in which the system is
enclosed.  In Eqs.~(\ref{correctiondef}) and (\ref{Z_eps2}), $\beta_1$
and $k_{00}$ are numerical constants which depend on the geometry of
the box considered~\cite{Hasenfratz,Hansen90}. Together with
Eqs.~(\ref{Sigeff}) and (\ref{Feff}), these equations imply:
\bea
\Sigma_{ef\!f}&=& \Sigma
\left( 1+\beta_1\frac{(N_f^2-1)}{N_fF^2\sqrt{V}}\right),
\nn\\
F_{ef\!f}&=& F
\left(1+\left[\beta_1+\frac{T^2}{\sqrt{V}}k_{00}\right]\frac{N_f}{2F^2\sqrt{V}}\right),
\nn
\eea
In the particular case of hypercube, i.e. a box with sides
$T{=}L_1{=}L_2{=}L_3{=}V^{1/4}$, $k_{00}=\beta_1/2$.

\subsection{Two-point correlation functions}

We consider here two-point correlators of the scalar and pseudoscalar quark
bilinears, 
\bea
S_0(x)&\equiv& \bar{\psi}(x)\one_{N_f}\psi(x)\ ,\ \ 
S_b(x)\equiv \bar{\psi}(x)t_b\one_{N_f}\psi(x)\ ,\nn\\ 
P_0(x)&\equiv& \bar{\psi}(x)i\gamma_5\one_{N_f}\psi(x)\ ,\ \ 
P_b(x)\equiv\bar{\psi}(x)t_b i\gamma_5\one_{N_f}\psi(x)\ , 
\label{bilinear}
\eea
where the $t_a$ denote the $SU(N_f)$
generators, and we have normalised $\Tr[(t_a)^2]=\frac12$. 
For $N_f=2$ we have $t_a=\frac12\sigma_a$ 
in Eq.~(\ref{bilinear}),
the Pauli matrices for $a=1,2,3$, and $\frac12$ times the identity for 
$t_{a=0}$.

In the effective theory, the scalar and pseudoscalar densities are
most easily obtained by introducing Hermitean sources, $s(x)=s_a(x)t_a$ and
$p(x)=p_a(x)t_a$, which have the same spurionic transformation properties
as in QCD, i.e. through the replacement~\cite{Gasser:1984gg}:
\be
\mathcal{M}\to \mathcal{M}+s(x)+i\, p(x)
\ .
\ee
The $\eps$-expansion counting for the sources is thus the same as that
of the quark masses, i.e. $s(x),\,p(x)\sim\eps^2$.

Once this replacement is made, the partition function depends on the
sources, and the two-point functions are obtained, as usual, by taking
adequate functional derivatives:
\be
\langle S_a(x) S_b(0) \rangle_{U(x)}\ = \  
\frac{1}{\cal Z}\,
\frac{\delta^2}{\delta s_a(x)\delta s_b(0)}
{\cal Z}[s,p]\Big|_{s=p=0}
\label{SSdef}
\ee
and likewise for the pseudoscalar correlators. Expanding the
observables, the action and the Jacobian to $\mathcal{O}(\eps^2)$, a
calculation analogous to the one performed for the partition function in
the preceeding section yields:
\bea
\langle S_0(x) S_0(0) \rangle_{U(x)} &=&\frac{\Sigma_{ef\!f}^2}{4}
\left\langle (\Tr[U_0+U_0^\dag])^2 \right\rangle_{U_0}
-\frac{\Sigma_{ef\!f}^2}{2F^2} 
\Big[ \langle \Tr\left[(U_0-U_0^\dag)^2\right]\rangle_{U_0}
-\frac{1}{N_f} \left\langle (\Tr[U_0-U_0^\dag])^2\right\rangle_{U_0}
\Big]\bD(x)\nn\\ 
&+&
\mathcal{O}(\eps^3),
\label{SSresult}
\eea
where all that remains are expectation values with respect to the zero
mode, given by:
\be
\langle O \rangle_{U_0} \equiv
\frac{1}{\int\! d_H U_0\ e^{-{\cal
  S}_{U_0}^{(0)}(\Sigma_{ef\!f},F_{ef\!f})}}
\int_{SU(N_f)} d_H U_0\ O\ 
e^{-{\cal
  S}_{U_0}^{(0)}(\Sigma_{ef\!f},F_{ef\!f})}
\ .\label{U0def}
\ee
In the second term of Eq.~(\ref{SSresult}), since $\bD(x)$ is
$\mathcal{O}(\eps^2)$, the expectation values with respect to $U_0$
can be calculated with ${\cal S}_{U_0}^{(0)}(\Sigma,F)$ instead of
${\cal S}_{U_0}^{(0)}(\Sigma_{ef\!f},F_{ef\!f})$.

Along the same lines, one can derive a 
similar expression 
for the pseudoscalar correlator:
\bea
\langle P_0(x) P_0(0) \rangle_{U(x)} \!
&=&\!\!-\frac{\Sigma_{ef\!f}^2}{4}
\left\langle (\Tr[U_0-U_0^\dag])^2 \right\rangle_{U_0}
+\frac{\Sigma_{ef\!f}^2}{2F^2} 
\Big[ \langle \Tr\left[(U_0+U_0^\dag)^2\right]\rangle_{U_0} 
-\frac{1}{N_f} \left\langle (\Tr[U_0+U_0^\dag])^2\right\rangle_{U_0}
\Big]\bD(x)
\nn\\
&+&
\mathcal{O}(\eps^3)
\ .\label{PPresult}
\eea
Both results Eqs.~(\ref{SSresult}) and  (\ref{PPresult}) 
have the same form as the corresponding 
expressions at zero chemical potential, $\mu=0$ \cite{DDHJ}. However,  
when inserting 
the group integral averages over $U_0$ below, they will differ explicitly 
by $\mu$-dependent terms.

Flavoured two-point functions can be computed exactly in the same way. 
But there is an important difference to the calculation in \cite{DDHJ}: 
while for $\mu=0$ and
equal quark masses one has $\langle S_a(x) S_b(0)
\rangle_{U(x)}\sim\delta_{ab}$, this is explicitly broken by the chemical
potential $\mu\neq0$ (as well as by non-degenerate masses of course). 

For that reason we will only compute the following diagonal 
sum over flavoured combinations, 
where we can use the $SU(N_f)$ completeness relation 
$\sum_a (t_a)_{ij}(t_a)_{kl}\ =\ \frac12 (\delta_{il}\delta_{jk}
-\frac{1}{N_f}\delta_{ij}\delta_{kl})$:
\bea
\sum_a\langle S_a(x) S_a(0)\rangle_{U(x)}
&=&\frac18\Sigma_{ef\!f}^2\left[ 
\langle \Tr\left[(U_0+U_0^\dag)^2\right]\rangle_{U_0}
-\frac{1}{N_f} \left\langle (\Tr[U_0+U_0^\dag])^2\right\rangle_{U_0}
\right]\nn\\
&&+\ \bD(x)\frac{\Sigma_{ef\!f}^2}{4F^2}\left[
-\frac12 \left\langle (\Tr[U_0+U_0^\dag])^2\right\rangle_{U_0}
-\frac{1}{2N_f^2}(N_f^2+2) \left\langle (\Tr[U_0-U_0^\dag])^2\right
\rangle_{U_0}
\right.\nn\\
&&\left.\ \ \ \ \ \ \ \ \ \ \ \ \ \ \ \ \ \ 
+2N_f^2 +\frac{2}{N_f} \langle \Tr\left[(U_0-U_0^\dag)^2\right]\rangle_{U_0}
\right]+\mathcal{O}(\eps^3)\ .
\label{SaSa}
\eea
For clarity we have made the sum over flavour indices explicit (we always use
summation conventions unless otherwise stated).
In complete analogy, one obtains for the flavoured pseudoscalar correlation
function,
\bea
\sum_a\langle P_a(x) P_a(0)
\rangle_{U(x)}
&=&-\ \frac18\Sigma_{ef\!f}^2\left[ 
\langle \Tr\left[(U_0-U_0^\dag)^2\right]\rangle_{U_0}
-\frac{1}{N_f} \left\langle (\Tr[U_0-U_0^\dag])^2\right\rangle_{U_0}
\right]\nn\\
&&-\bD(x)\frac{\Sigma_{ef\!f}^2}{4F^2}\left[
-\frac12 \left\langle (\Tr[U_0-U_0^\dag])^2\right\rangle_{U_0}
-\frac{1}{2N_f^2}(N_f^2+2) \left\langle (\Tr[U_0+U_0^\dag])^2\right
\rangle_{U_0}
\right.\nn\\
&&\left.\ \ \ \ \ \ \ \ \ \ \ \ \ \ \ \ \ 
-2N_f^2 +\frac{2}{N_f} \langle \Tr\left[(U_0+U_0^\dag)^2\right]\rangle_{U_0}
\right]+\mathcal{O}(\eps^3)\ .
\label{PaPa}
\eea
Both results agree again with \cite{DDHJ} at $\mu=0$, when normalised
by $(N_f^2-1)$. The resulting group averages differ however, both by
explicitly $\mu$-dependent factors and by a functional change of the mass
dependent condensate.  This is the subject of the next section.

Before turning to the evaluation of the relevant group integrals, it
is useful to make a comment on the spacetime dependence of the above
correlation functions. For comparisons with lattice QCD calculations,
it is useful to consider the corresponding zero-momentum correlation
functions, which are functions only of the Euclidean time $t$, i.e.
\be
C_S(t)\equiv \frac{T}{V}\int d^3x\,\langle S_a(x) S_a(0)
\rangle_{U(x)}
\qquad\mbox{and}\qquad C_P(t)\equiv  \frac{T}{V}\int d^3x\,\langle P_a(x)
P_a(0) 
\rangle_{U(x)}
\ ,
\label{eq:zermomcorrs}\ee
where either $a=0$, for the singlet case, or there is an implicit sum
over the $SU(N_f)$ adjoint index $a$, for the flavoured case. As above, in
Eq.~(\ref{eq:zermomcorrs}) $T$ is the time extent of the box in which
the system is enclosed and $V/T$ is its spatial volume. For the
massless propagtor~\cite{Hasenfratz},
\be
\bar{\Delta}(x) \equiv \frac{1}{V}\sum_p\!'
~\frac{e^{ip\cdot x}}{p^2}\qquad\Rightarrow\qquad 
h_1(\tau) \equiv
\frac{1}{T}\int\! d^3x\, \bar{\Delta}(x)=
\frac{1}{2}\left[(\tau - \frac{1}{2})^2 - \frac{1}{12}\right]
\ ,\ee
with $0<\tau<1$ and where $\tau\equiv t/T$. Thus, the zero-momentum
correlation functions $C_{S,P}(t)$ are simply obtained from the
expressions of Eqs.~(\ref{SSresult}), (\ref{PPresult}), (\ref{SaSa})
and (\ref{PaPa}) above, and those of Eqs.~(\ref{SSfinal}),
(\ref{PPfinal}), (\ref{Safinal}) and (\ref{Pafinal}) below, by making
the replacement:
\be
\bar{\Delta}(x) \longrightarrow \frac{T^2}{V}\, h_1(\tau)
\ .
\ee
%

\sect{Results for scalar and pseudoscalar correlators
at fixed topology}
\label{Results}

In order to make the group integrals, which appear in the results of
the previous section, tractable analytically, we choose to work in
sectors of fixed gauge-field topology. This is done by introducing the
theta vacuum angle, $\theta$, of QCD into the effective theory, through
the replacement~\cite{Gasser:1984gg}
\be
\mathcal{M}+s(x)+i\,p(x)\to \big(\mathcal{M}+s(x)+i\,p(x)\big)\,e^{i\theta/N_f}
\ee
and Fourier transforming the results with respect to this angle. Thus
for instance, the partition function in the sector of topology $\nu$
is given by:
\be
\mathcal{Z}_\nu\equiv
\int_0^{2\pi}\frac{d\theta}{2\pi}\,e^{-i\theta\nu}\,\mathcal Z[\theta] 
\ .
\ee
It is easy to convince oneself that such a transformation has for
effect to replace all $SU(N_f)$ group integrals by integrals over the
much simpler group manifold of $U(N_f)$. Thus, to $\mathcal{O}(\eps^2)$:
\bea
{\cal Z}_{\nu}&=& \int_{U(N_f)} d_H U_0 \det[ U_0^\nu]
\exp\left[\frac12VF^2_{ef\!f}\mu^2\Tr[\Ga U_0^\dag \Ga U_0 ]
+ \frac12 V\Sigma_{ef\!f}m\Tr[U_0+U_0^\dag]\right]
+\mathcal{O}(\eps^3)\ ,
\label{Znu}\\
\langle\ O\ \rangle_{U_0}^\nu &\equiv&
\frac{1}{{\cal Z}_{\nu}}
\int_{U(N_f)} d_H U_0\, O\det[U_0^\nu]
\exp\left[\frac12 VF^2_{ef\!f}\mu^2\Tr[\Ga U_0^\dag \Ga U_0 ] 
+ \frac12 V\Sigma_{ef\!f}m\Tr[ U_0+U_0^\dag]\right]
\!+\mathcal{O}(\eps^3)
\ ,\nn\\
\label{U0vev}
\eea
where we have absorbed the constant $\mathcal{N}$ of Eq.~(\ref{eq:Zeps2}) 
into the measure $d_H U_0$,
as well as the constant term $\frac14 VF^2_{ef\!f}\mu^2\Tr[\Ga^2]$ from 
${\cal  S}_{U_0}^{(0)}(\Sigma_{ef\!f},F_{ef\!f})$.

In the following, we only consider the theory with equal masses, 
\be
{\cal M}\ =\ m\one_{N_f}\ . 
\ee
This
includes the interesting case of $N_f=2$ with degenerate up
and down quarks. 
Furthermore, we specify the external vector current to be \be C\ =\
\mu\Ga\ ,\ \ \mbox{with} \ \ \Ga\ \equiv\
\mbox{diag}(\one_{N_1},-\one_{N_2})\ , \ee such that
$\Ga^2=\one_{N_f}$ and we will also consider more general partition
functions, containing higher powers $ \Tr[(\Ga U_0^\dag \Ga U_0)^k ]$
as generating functionals, for computing all group integrals.

Correlation functions at fixed topology are obtained by
replacing the averages 
$\langle\ O\ \rangle_{U_0}\to\langle\ O\ \rangle_{U_0}^\nu$
in Eqs.~(\ref{SSresult}) and (\ref{PPresult}). 
The building blocks that we have to compute in
Eqs.~(\ref{SSresult}) and  (\ref{PPresult}) are 
$ \langle(\Tr [U_0])^2+ (\Tr [U_0^\dag])^2\rangle_{U_0}^\nu$, 
$ \langle\Tr [U_0]\Tr [U_0^\dag]\rangle_{U_0}^\nu$, and 
$ \langle\Tr[ U_0^2]+ \Tr[ U_0^{\dag\,2}]\rangle_{U_0}^\nu$.
With the help of some $U(N_f)$ group integral identities derived in 
Appendix \ref{B}, these can be expressed
in terms of the following known quantities: the partition function and its
derivatives with respect to mass, chemical potential and an additional
external field. The equations that we need are: 
\bea
\left\langle (\Tr[U_0+U_0^\dag])^2\right\rangle_{U_0}^\nu&=& 
4N_f
\frac{\partial}{\partial \eta_{ef\!f} }
\frac{\Sigma_\nu(\eta_{ef\!f},\alpha_{ef\!f})}{\Sigma}
+4N_f^2\frac{\Sigma_\nu(\eta_{ef\!f},\alpha_{ef\!f})^2}{\Sigma^2} \ ,
\label{TrU+U2}\\
\left\langle (\Tr[U_0-U_0^\dag])^2\right\rangle_{U_0}^\nu&=& 
-4\frac{N_f}{\eta_{ef\!f}}
\frac{\Sigma_\nu(\eta_{ef\!f},\alpha_{ef\!f})}{\Sigma}+
\frac{4\nu^2N_f^2}{\eta_{ef\!f}^2}\ ,
\label{TrU-U2}
\eea
for the squared traces, and for single traces of the squared matrices
\bea
\langle \Tr\left[(U_0\pm U_0^\dag)^2\right]\rangle_{U_0}^\nu
&=& 
-4\frac{N_f^2}{\eta_{ef\!f}}
\frac{\Sigma_\nu(\eta_{ef\!f},\alpha_{ef\!f})}{\Sigma}+
\frac{4\nu^2N_f}{\eta_{ef\!f}^2}
+(2\pm2)N_f
\nn\\
&&+\frac{4\al_{ef\!f}^2}{\eta_{ef\!f}^2}
\Big(N_f{\cal Y}_\nu(\eta_{ef\!f},\alpha_{ef\!f}) -(\Tr[\Gamma])^2+
\frac12\al_{ef\!f}^2({\cal X}_\nu(\eta_{ef\!f},\alpha_{ef\!f})-N_f)\Big)\ .
\label{TrU2}
\eea
The generating functionals on the right hand sides are as follows. The
mass dependent condensate is given by
\be
\frac{\Sigma_\nu(\eta_{ef\!f},\alpha_{ef\!f})}{\Sigma}\ \equiv\ 
\frac{1}{N_f} \frac{\partial}{\partial \eta_{ef\!f}}\ln [{\cal
  Z}_{\nu}]\ =\ \frac{1}{2N_f}
\langle\Tr[ U_0+ U_0^{\dag}]\rangle_{U_0}^\nu\ .
\label{resolvent}
\ee
It depends both on the 
rescaled mass 
\be
\eta_{ef\!f}\ \equiv\  m\Sigma_{ef\!f}V \ ,
\label{eta}
\ee
and the rescaled
chemical potential,
\be 
\alpha^2_{ef\!f}\ \equiv\ \mu^2F^2_{ef\!f}V \ .
\label{alf}
\ee
We will also need the derivative of the condensate to generate the
expectation value $\langle(\Tr[ U_0+
U_0^{\dag}])^2\rangle_{U_0}^\nu$.  Whether or not to consider the
${\cal O}(\eps^2)$ correction is trivially achieved by keeping or
dropping the subscript ``$ef\!f$'' in the couplings in
Eqs.~(\ref{resolvent}) and (\ref{Y}).  For example, the partition
function ${\cal Z}_{\nu}$ to leading order is given by Eq.~(\ref{Znu})
with leading order couplings $\Sigma$ and $F$.

We define a second functional, which is the derivative of the
partition function with respect to the rescaled chemical potential, to
generate the following average: 
\be {\cal
  Y}_\nu(\eta_{ef\!f},\alpha_{ef\!f})\ \equiv\
2\frac{\partial}{\partial \alpha^2_{ef\!f}}\ln [{\cal Z}_{\nu}]\ =\
\langle\Tr[ \Ga U_0^{\dag}\Ga U_0]\rangle_{U_0}^\nu\ .
\label{Y}
\ee
We will also need a third generating functional 
\be
{\cal X}_\nu(\eta_{ef\!f},\alpha_{ef\!f})\ \equiv\ 
\langle\Tr\left[ (\Ga U_0^{\dag}\Ga U_0)^2\right]\rangle_{U_0}^\nu\ ,
\label{X}
\ee
to be given below in Eq.~(\ref{Xresult}) in order to
complete the calculation.

\subsection{Unflavoured correlation functions}

We begin with the scalar two-point function, inserting Eqs.~(\ref{TrU+U2}) -
(\ref{TrU2}) into Eq.~(\ref{SaSa}) to obtain
\bea
\langle S_0(x) S_0(0) \rangle_{U(x)}^\nu
&=&{\Sigma^2_{ef\!f}N_f}\left(
\frac{\partial}{\partial \eta_{ef\!f}}
\frac{\Sigma_\nu(\eta_{ef\!f},\alpha_{ef\!f})}{\Sigma}
+{N_f}\frac{\Sigma_\nu(\eta_{ef\!f},\alpha_{ef\!f})^2}{\Sigma^2}\right) 
\nn\\
&+&
\bD(x)\frac{2\Sigma^2}{\eta^2 F^2}
\left[ (N_f^2-1) \eta\frac{\Sigma_\nu(\eta,\alpha)}{\Sigma}
-{\alpha^2}
\Big(N_f{\cal Y}_\nu(\eta,\alpha) -(\Tr[\Gamma])^2
+\frac{\alpha^2}{2}({\cal X}_\nu(\eta,\alpha)-N_f)
\Big)\right]\nn\\
&+& {\cal O}(\eps^3)\ .
\label{SSfinal}
\eea Here $\langle\cdots\rangle_{U(x)}^\nu$ denotes the expectation
value with respect to the full field $U(x)$ at fixed topology $\nu$.
In the term proportional to $\bD(x)$, the subscripts $ef\!f$ can be
dropped since $\bD(x)$ is already of $\mathcal{O}(\eps^2)$.  The form
of the first three terms is identical to the result for $\mu=0$ in
\cite{DDHJ}. The only difference is that here, the resolvent
$\Sigma_\nu(\eta,\alpha)$ also depends on the rescaled chemical
potential. The remaining term proportional to
$\alpha^2$ is an explicit correction to the result of \cite{DDHJ}, at
non-vanishing chemical potential.

For the pseudoscalar two-point function, we obtain in the same way
\bea
\langle P_0(x) P_0(0) \rangle_{U(x)}^\nu
&=& \frac{\Sigma^2_{ef\!f}N_f}{\eta_{ef\!f}}\left(
\frac{\Sigma_\nu(\eta_{ef\!f},\alpha_{ef\!f})}{\Sigma}
-\frac{\nu^2N_f}{\eta_{ef\!f}}\right)
\nn\\
&+&\bD(x)\frac{2\Sigma^2}{F^2}
\left[ 
-\frac{\partial}{\partial \eta}\frac{\Sigma_\nu(\eta,\alpha)}{\Sigma}
-N_f\frac{\Sigma_\nu(\eta,\alpha)^2}{\Sigma^2}
-\frac{N_f^2}{\eta}\frac{\Sigma_\nu(\eta,\alpha)}{\Sigma}
+\frac{\nu^2N_f}{\eta^2}
+N_f
\right.\nn\\
&&\left.
\ \ \ \ \ \ \ \ \ \ \ \ \ +\frac{\alpha^2}{\eta^2}
\Big(N_f{\cal Y}_\nu(\eta,\alpha)-(\Tr[\Ga])^2 
+\frac{\alpha^2}{2}({\cal X}_\nu(\eta,\alpha)-N_f)
\Big)
\right]\ + {\cal O}(\eps^3)\ .
\label{PPfinal}
\eea
As before, all terms but the last are of the same form as for $\mu=0$,
but now depend on the rescaled chemical potential $\al$ and thus $F$,
also at leading order. The term in the last line proportional to
$\alpha^2$ is again an explicit correction term.

We now explicitly give the three functions appearing in the results for the
two-point correlations above in the particular case of $N_f=2$ flavours. 
The results for more flavours follow easily using Appendix \ref{C}.

The $N_f=2$ flavour partition function 
reads \cite{SV04,AFV}
\be
\underline{N_f=2}:\ \ 
{\cal Z}_{\nu}(\eta_{ef\!f}) 
= \int_0^1d\la \la\ e^{\frac12\al_{ef\!f}^2(4\la^2-2)} 
I_\nu(\la\,\eta_{ef\!f})^2 \ ,
\label{ZNf2}
\ee
where $I_\nu$ denotes the modified $I$-Bessel function. 

From the above equations, the condensate of Eq.~(\ref{resolvent})
easily follows, and for two flavours we arrive at 
\be
\underline{N_f=2:}\ \
\frac{\Sigma_\nu(\eta_{ef\!f},\alpha_{ef\!f})}{\Sigma}\ =\
\frac{1}{{\cal Z}_{\nu}(\eta_{ef\!f})} \int_0^1 d\la \la\
e^{\frac12\al_{ef\!f}^2(4\la^2-2)} I_\nu(\la\,\eta_{ef\!f})\Big(
\frac{\nu}{\eta_{ef\!f}}I_\nu(\la\,\eta_{ef\!f}) +\la
I_{\nu+1}(\la\,\eta_{ef\!f})\Big).
\label{SigmaNf2}
\ee
The second quantity ${\cal Y}_\nu(\eta,\al)$ 
equally follows from
its definition (\ref{Y}) by differentiation.
Using Eq.~(\ref{ZNf2}) we obtain  for two flavours 
\be
\underline{N_f=2:}\ \ {\cal Y}_\nu(\eta_{ef\!f},\alpha_{ef\!f})\ =\  
\frac{2}{{\cal    Z}_{\nu}(\eta_{ef\!f})}
\int_0^1 d\la\la \frac12(4\la^2-2)\, e^{\frac12\al_{ef\!f}^2(4\la^2-2)} 
I_\nu(\la\,\eta_{ef\!f})^2 \ ,
\label{calNNf2}
\ee
given here to ${\cal O}(\eps^2)$, 
which we will need for the flavoured correlators below.

The third function is obtained by differentiation of the following
generalised partition function computed in Appendix \ref{C} 
\be
{\cal Z}_{gen}
\ \equiv\ \int_{U(N_f)} d_H U_0 \det[ U_0^\nu] 
e^{\,\frac12V F_{ef\!f}^2\mu^2 \Tr[\Ga U_0^\dag \Ga U_0 ]
+ \frac12V\Sigma_{ef\!f}m\Tr[U_0+U_0^\dag]\ +\ 
\omega \Tr\Big[(\Ga U_0^\dag \Ga U_0)^2 \Big]}\ .
\label{ZNfbeta}
\ee 
It is remarkable that this group integral can be calculated, even
when adding any power $\Tr\Big[(\Ga U_0^\dag\Ga U_0)^k\Big]$ to the
partition function of Eq.~(\ref{Znu}), as was pointed out in
\cite{JacLH}.  We generalise the result given in \cite{JacLH} in
Appendix \ref{C} to arbitrary masses ${\cal M}\neq m\one_{N_f}$, with
$\Tr[\Ga]\neq0$. For two flavours this generalised partition function
reads 
\be {\cal Z}_{gen}(\eta_{ef\!f}) = \int_0^1d\la \la\
e^{\frac12\al_{ef\!f}^2(4\la^2-2) +\omega\big(16(\la^4-\la^2)+2\big)}
I_\nu(\la\,\eta_{ef\!f})^2
\label{ZNf2beta}
\ee
and for more flavours the result is given by 
Eq.~(\ref{Zgenresult}). We then obtain
\be
{\cal X}_\nu(\eta_{ef\!f},\alpha_{ef\!f})\ =\ 
\frac{\partial}{\partial \omega}
\ln [{\cal  Z}_{gen}]\Big|_{\omega=0}\ =\ 
\langle\Tr\left[ (\Ga U_0^{\dag}\Ga U_0)^2\right]\rangle_{U_0}^\nu \ ,
\label{Xresult}
\ee
making only the ordinary partition function 
of Eq.~(\ref{ZNf2}) appear in the denominator. 
We again display the two-flavour result at equal mass,
\be
\underline{N_f=2:}\ \ {\cal X}_\nu(\eta_{ef\!f},\alpha_{ef\!f})\ =\ 
\frac{1}{{\cal Z}_{\nu}(\eta_{ef\!f})}  
\int_0^1d\la\la\,\big(16(\la^4-\la^2)+2\big)\,
e^{\frac12\al_{ef\!f}^2(4\la^2-2)} I_\nu(\la\,\eta_{ef\!f})^2\ .
\ee
%

\subsection{Flavoured correlation functions}

We now give the sums over flavoured two-point correlations, which are
easier to compute on the lattice, since they do not involve
disconnected contributions. This is just a matter of inserting
Eqs.~(\ref{TrU+U2}) - (\ref{TrU2}) into the results of Eqs.~(\ref{SaSa})
and (\ref{PaPa}) at fixed topology. Here it is useful to observe that
we have already computed the combinations $\langle \Tr\left[(U_0\pm
  U_0^\dag)^2\right]\rangle_{U_0} -\frac{1}{N_f} \left\langle
  (\Tr[U_0\pm U_0^\dag])^2\right\rangle_{U_0}$ in the unflavoured
cases. We obtain for the scalar two-point function, 
\bea \sum_a\langle S_a(x)
S_a(0)\rangle_{U(x)}^\nu&=& \frac{\Sigma_{ef\!f}^2}{2}\left[
  -\frac{\partial}{\partial \eta_{ef\!f}}
  \frac{\Sigma_\nu(\eta_{ef\!f},\alpha_{ef\!f})}{\Sigma}
  -N_f\frac{\Sigma_\nu(\eta_{ef\!f},\alpha_{ef\!f})^2}{\Sigma^2}
  -\frac{N_f^2}{\eta_{ef\!f}}
  \frac{\Sigma_\nu(\eta_{ef\!f},\alpha_{ef\!f})}{\Sigma}
  +\frac{\nu^2N_f}{\eta_{ef\!f}^2}
\right.\nn\\
&&\ \ \ \ \ \ \ \ \ +
N_f+\left.\frac{\alpha_{ef\!f}^2}{\eta_{ef\!f}^2} \Big(N_f{\cal
    Y}_\nu(\eta_{ef\!f},\alpha_{ef\!f}) -(\Tr[\Ga])^2
  +\frac{\alpha_{ef\!f}^2}{2}({\cal
    X}_\nu(\eta_{ef\!f},\alpha_{ef\!f})-N_f) \Big) \right]
\nn\\
&+&\bD(x)\frac{\Sigma^2}{2F^2}\left[
  -N_f\frac{\partial}{\partial\eta}\frac{\Sigma_\nu(\eta,\alpha)}{\Sigma}
  -N_f^2\frac{\Sigma_\nu(\eta,\alpha)^2}{\Sigma^2}
  +\frac{(-3N_f^2+2)}{N_f\eta}\frac{\Sigma_\nu(\eta,\alpha)}{\Sigma}
  -(N_f^2-2)\frac{\nu^2}{\eta^2}
\right.\nn\\
&&\left.  \ \ \ \ \ \ \ \ \ \ \ \ \ \
  +N_f^2+\frac{4\al^2}{N_f\eta^2}\Big(N_f{\cal Y}_\nu(\eta,\alpha)
  -(\Tr[\Ga])^2 +\frac{\alpha^2}{2}({\cal
    X}_\nu(\eta,\alpha)-N_f)\Big)
\right]\ + {\cal O}(\eps^3)\ ,\nn\\
\label{Safinal}
\eea
and for the pseudoscalar correlator,
\bea
\sum_a\langle P_a(x) P_a(0)\rangle_{U(x)}^\nu\!\!&=&\!\!
\frac{\Sigma_{ef\!f}^2}{2}\!\left[
\frac{(N_f^2-1)}{\eta_{ef\!f}}
\frac{\Sigma_\nu(\eta_{ef\!f},\alpha_{ef\!f})}{\Sigma}
\right.\nn\\
&&\left.\ \ \ \ \ \ \ \ -\frac{\alpha_{ef\!f}^2}{\eta_{ef\!f}^2}
\Big(N_f{\cal Y}_\nu(\eta_{ef\!f},\alpha_{ef\!f})-(\Tr[\Ga])^2 
+\frac{\alpha_{ef\!f}^2}{2}({\cal X}_\nu(\eta_{ef\!f},\alpha_{ef\!f})-N_f)
\Big)\right]\nn\\
&+&\bD(x)\frac{\Sigma^2}{2F^2}\left[
\frac{(N_f^2+2)}{N_f}\Big( 
\frac{\partial}{\partial\eta}\frac{\Sigma_\nu(\eta,\alpha)}{\Sigma}
+ N_f\frac{\Sigma_\nu(\eta,\alpha)^2}{\Sigma^2}\Big)
+3\frac{N_f}{\eta}
\frac{\Sigma_\nu(\eta,\alpha)}{\Sigma}
+(N_f^2-4)\frac{\nu^2}{\eta^2}
\right.
\nn\\
&&\left.\ \ \ \ \ \ \ \ \ \ \ \ \ \  +N_f^2 -4
-\frac{4\al^2}{N_f\eta^2}\Big(N_f{\cal Y}_\nu(\eta,\alpha)-(\Tr[\Ga])^2  
+\frac{\alpha^2}{2}({\cal X}_\nu(\eta,\alpha)-N_f)\Big)
\right]\nn\\ 
&+& {\cal O}(\eps^3)\ ,
\label{Pafinal}
\eea
where we now have explicit 
$\mu$-dependent corrections at leading order, which is not the case for the
other correlation functions.

\section{Conclusions}
\label{conc}
We have computed scalar and pseudoscalar two-point correlation
functions in the epsilon regime of Chiral Perturbation Theory, up to
and including corrections of ${\cal O}(\eps^2)$. The main new feature
of our results is the inclusion of an isospin chemical potential of
real or imaginary type. This leads to the appearance of the low energy
constant $F$ already at leading ${\cal O}(\eps^0)$. We find
corrections to the meson correlation functions obtained at $\mu=0$,
which contain both explicit terms proportional to $\al^2=\mu^2F^2V$
and $\al^4$, as well as implicit corrections which arise through those
to the partition function and its derivatives.  All of our results are
obtained for $N_f$-flavour QCD. Possible extensions of our work
include the calculation of correlations functions in quenched or
partially quenched QCD, which would require computing supersymmetric
extensions of the group integrals.  Axial or vector current two-point
functions are also feasible.

Our results provide alternate means of extracting the low energy
constants $\Sigma$ and $F$ from lattice calculations. In particular,
the additional parameter $\mu$ and the fact that $F$ appears at
leading order should help improve the precision in the determination
of this low energy constant. Moreover, we have obtained a new one-loop
result for $F_{ef\!f}$, which includes one of the
$\mathcal{O}(\eps^2)$ corrections to the partition function.

\indent

\noindent
\underline{Acknowledgements}:

We thank Poul Damgaard, Tom DeGrand and Hide Fukaya for discussions.
The hospitality of the CPT Luminy is gratefully acknowledged by
G.A. and F.B., where part of this work was performed.  This work was
supported in part by the EU networks ENRAGE MRTN-CT-2004-005616,
FLAVIAnet MRTN-CT-2006-035482, by EPSRC grant EP/D031613/1 and by the
CNRS's GDR grant $n^o$ 2921 (“Physique subatomique et calculs sur
r\'eseau”).

\appendix

\sect{
Derivation of zero-mode group integral identities for $\mu\neq0$
}\label{B}

In this appendix we will derive a set of $U(N_f)$ group identities
among expectation values of averages over various traces.  Those
relations are needed to arrive at Eqs.~(\ref{TrU+U2}) - (\ref{TrU2})
in the expressions for the two-point functions.

In particular we have to specify here
\be
C\ =\ \mu\Gamma\ = \ \mu\ \mbox{diag}(\one_{N_1},-\one_{N_2})\ ,
\label{gamma}
\ee
with $N_1\neq N_2$ in general. Furthermore let us stress that our
identities hold for degenerate masses only,
\be
V\Sigma\mathcal{M}\ =\ \eta\,\one_{N_f}
\ee
where $\eta$ is the rescaled mass. 
In order to simplify notation we will use
\bea
{\cal Z}_{\nu}&\equiv& \int_{U(N_f)} d_H U \det[ U^\nu]
\exp\left[\frac12\al^2\Tr[\Gamma U^\dag \Gamma U ]
+ \frac12\eta\Tr[U+U^\dag]\right],
\label{Z}\\
\langle\ O\ \rangle&\equiv&
\frac{1}{{\cal Z}_{\nu}}
\int_{U(N_f)} d_H U\ O\ \det[U^\nu]\exp\left[
\frac{1}{2}\al^2\Tr[\Gamma U^\dag \Gamma U ]+\frac{1}{2}\eta\Tr[U+U^\dag]
]\right],
\label{vev}
\eea
dropping most indices from the main text. In particular here $U=U_0$ denotes
the constant $U(N_f)$ matrix.
The full ${\cal O}(\eps^2)$ improved expectation values Eqs.~(\ref{Znu})
and (\ref{U0vev}) are trivially obtained by shifting $\al\to\al_{ef\!f}$ and 
$\eta\to\eta_{ef\!f}$.

Following \cite{DDHJ} we introduce the explicit representation of
the left differentiation with respect to group elements $U_{kl}$ of $U(N_f)$
\be
\nabla_a\ \equiv\ i(t_aU)_{kl}\frac{\partial}{\partial U_{kl}} \ .
\label{DL}
\ee
Throughout this appendix, the $t_a$ denote the generators of the algebra of
$u(N_f)$. They satisfy the $U(N_f)$ completeness relation 
\be
(t_a)_{ij}(t_a)_{kl}\ =\ \frac12 \delta_{il}\delta_{jk}
\ee
in the normalisation $\Tr[t_at_b]=\frac12\delta_{ab}$.
For example, we obtain
\bea
\nabla_a U &=& +it_aU\ ,\ \ \nabla_a U^\dag\ =\ -iU^\dag t_a\ ,
\label{DU}\\
\nabla_a \det[U] &=& i \Tr[t_a]\det[U]\ .\nn
\eea
Due to the left invariance of the Haar measure, integrals over total
derivatives with respect to $\nabla_a$ vanish:
\be
0\ =\ \int_{U(N_f)} d_H U\ \nabla_a\Tr\left[t_aG(U) \det[ U^\nu]
\exp\left[\frac12\al^2\Tr[\Gamma U^\dag \Gamma U ]
+ \frac12\eta\Tr[U+U^\dag]\right]
\right],
\label{startid}
\ee
for any function $G(U)$. By choosing a suitable set of functions, we will
generate a closed set of equations for averages that can be solved for in
terms of known generating functions. 

The simplest choice in Eq.~(\ref{startid}) is $G(U)=\one$ and $a=0$, without
summing over $a$, which leads to the
following identity
\be
0\ =\ \nu N_f 
+ \frac{\eta}{2}\langle \Tr [U-U^\dag] \rangle\ .
\label{III}
\ee 
This equation is invariant under complex conjugation when simultaneously
changing $\nu\to-\nu$ (see the measure in Eq.~(\ref{vev})). 
Next, in choosing $G(U)=U-U^\dag$ and keeping  $a=0$, we obtain:
\be
0 = \frac{\eta}{2}\Big\langle (\Tr[ U-U^\dag])^2\Big\rangle 
+\nu N_f \langle \Tr [U-U^\dag] \rangle
+\langle \Tr [U+U^\dag] \rangle\ . 
\label{VI}
\ee

Next, we sum over $a$ in Eq.~(\ref{startid})
is $G(U)=U$, leading to: 
\be
0\ =\ (N_f+\nu)\langle \Tr[U] \rangle 
+ \frac{\eta}{2}\left(\langle \Tr \Big[U^2\Big]\rangle-N_f\right)
+ \frac{\al^2}{2}
\left(\langle \Tr \Big[U^2\Gamma U^\dag \Gamma\Big] \rangle
- \langle \Tr[U]\rangle\right)\ .
\label{I}
\ee 
Either by  complex conjugation and changing $\nu\to-\nu$, or simply choosing 
$G(U)=-U^\dag$ we obtain
\be
0\ =\ (N_f-\nu)\langle \Tr[U^\dag] \rangle 
+ \frac{\eta}{2}\left(\langle \Tr \Big[U^{\dag\,2}\Big]\rangle-N_f\right)
+ \frac{\al^2}{2}\left(\langle \Tr \Big[U\Gamma U^{\dag\,2} \Gamma\Big] \rangle
- \langle \Tr[U^\dag]\rangle\right)\ .
\label{II}
\ee 
Taking the sum of Eqs.~(\ref{I}) and (\ref{II}) we obtain an equation
for $\langle \Tr \Big[(U-U^\dag)^2\Big]\rangle$.

The quantity $\langle \Tr \Big[U^2\Gamma U^\dag \Gamma\Big] \rangle$
and its conjugate which are new for $\al\neq0$ cannot be easily
derived from a known generating functional. Therefore we need an
additional set of equations compared to \cite{DDHJ}, which is
generated by choosing $G(U)=U\Gamma U^\dag\Gamma$
\be
0\ =\ 
\frac{\eta}{2}\left(\langle \Tr \Big[U^2\Gamma U^\dag \Gamma \Big]\rangle- 
\langle \Tr[U^\dag]\rangle\right)
+ (N_f+\nu)\langle \Tr [U\Gamma U^\dag \Gamma] \rangle
-\Big\langle(\Tr [\Gamma])^2\Big\rangle
+\frac{\al^2}{2}
\left(\langle \Tr\Big[( U\Gamma U^\dag \Gamma)^2\Big] \rangle-N_f\right) \ ,
\label{IV}
\ee
and its complex conjugate
\be
0\ =\ 
\frac{\eta}{2}\left(\langle \Tr \Big[U\Gamma U^{\dag\,2} \Gamma \Big]\rangle- 
\langle \Tr[U]\rangle\right)
+ (N_f-\nu)\langle \Tr[ U\Gamma U^\dag \Gamma] \rangle
-\Big\langle(\Tr [\Gamma])^2\Big\rangle
+\frac{\al^2}{2}
\left(\langle \Tr\Big[( U\Gamma U^\dag \Gamma)^2\Big] \rangle-N_f\right)\ .
\label{V}
\ee
We can now eliminate 
$\langle \Tr \Big[U^2\Gamma U^\dag \Gamma\Big] \rangle$ and its conjugate by
using the sum of Eqs.~(\ref{IV}) and (\ref{V}) to obtain
\bea
\langle \Tr \Big[(U-U^\dag)^2\Big]\rangle &=& 
-\frac{2N_f}{\eta}\langle \Tr [U+U^\dag] \rangle
-\frac{2\nu}{\eta}\langle \Tr [U-U^\dag] \rangle
\nn\\
&&+\frac{2\al^2}{\eta^2}\left[2N_f\langle \Tr[ U\Gamma U^\dag \Gamma] \rangle
-2\Big\langle(\Tr [\Gamma])^2\Big\rangle
+\al^2\Big( \langle \Tr\Big[( U\Gamma U^\dag \Gamma)^2\Big] \rangle-N_f\Big)
\right]\ .
\label{TrU2A}
\eea
All objects on the right hand side can now be generated. 
Differentiating the partition function with respect to the mass we have 
\be
\langle \Tr [U+U^\dag] \rangle\ =\ 2 \frac{\partial}{\partial\eta} 
\ln[{\cal Z}_{\nu}] \ = \ 2N_f\frac{\Sigma_\nu(\eta,\al)}{\Sigma}\ ,
\label{sigmaA}
\ee
using the definition Eq.~(\ref{resolvent}). Eq.~(\ref{III}) provides the
difference
\be
\langle \Tr [U-U^\dag] \rangle\ = -\frac{2\nu N_f}{\eta} \ .
\ee 
Differentiating the partition function with respect to $\al^2$ we obtain 
(see the definition Eq.~(\ref{Y}))
\be
\langle \Tr[ U\Gamma U^\dag \Gamma] \rangle\ =\ 
2 \frac{\partial}{\partial{\al^2}} \ln[{\cal Z}_{\nu}]\ =\ {\cal
  Y}_\nu(\eta,\al) \ .
\ee
The trace of the square is obtained from
the generalised partition function ${\cal Z}_{gen}$
derived in Appendix \ref{C} below, 
\be
\langle \Tr\Big[( U\Gamma U^\dag \Gamma)^2\Big] \rangle
\ =\ 
\frac{\partial}{\partial_\omega} \ln[{\cal Z}_{gen}]\Big|_{\omega=0}
\ =\ {\cal X}_\nu(\eta,\al)\ ,
\ee 
see the definition Eq.~(\ref{X}). Inserting all these into Eq.~(\ref{TrU2A}) 
we arrive at Eq.~(\ref{TrU2}), where 
$\langle \Tr \Big[(U+U^\dag)^2\Big]\rangle$ is obtained trivially  by adding 
$+4N_f$. 

The missing squares of traces in Eqs.~(\ref{SSresult}) and (\ref{PPresult})
follow  by differentiating twice with respect to the mass, 
\be
\Big\langle (\Tr[ U+U^\dag])^2\Big\rangle \ =\ 4\frac{1}{{\cal Z}_\nu}
\frac{\partial^2}{\partial_\eta^2} {\cal Z}_\nu
\ =\
4N_f\frac{\partial}{\partial\eta}\frac{\Sigma_\nu(\eta,\al)}{\Sigma}+4N_f^2 
\frac{\Sigma_\nu(\eta,\al)^2}{\Sigma^2}\ ,
\ee
leading to Eq.~(\ref{TrU+U2}).  Finally, a last equation is needed
that contains the square of the trace of the difference, which appears
in Eq.~(\ref{VI}) above. Using Eq.~(\ref{III}) and the generator of
Eq.~(\ref{sigmaA}), we obtain the Eq.~(\ref{TrU-U2}) in
Sec.~\ref{Results}, after replacing the couplings $\alpha$ and $\eta$
by their effective counterparts, $\alpha_{ef\!f}$ and $\eta_{ef\!f}$.


\sect{
Calculation of the generalised partition function
}\label{C}

In this appendix we compute the following generalisation of the
partition function of Eq.~(\ref{ZNfbeta})
\be 
{\cal Z}_{gen}(\{\eta\}) \ \equiv\ 
\int_{U(N_f)} d_HU_0 \det[U_0]^\nu 
\exp\left[{\frac12 \Tr[{\cal M} (U_0+U_0^\dag)]  
+\sum_p a_p\Tr\left[(U_0 \Ga U_0^\dag \Ga)^p\right]}\right] \ ,
\label{Zgen} 
\ee
where 
${\cal M}=\mbox{diag}(\eta_{f1=1},\cdots,\eta_{N_1},
\eta_{f2=1},\cdots,\eta_{N_2})$ contains the rescaled masses which may now
be different. 
As in the previous appendix, the inclusion of effective couplings is trivial. 
Without loss of generality we choose $N_{2}\geq N_1$ in $N_f=N_1+N_2$.
The volume $V$ and higher order coupling constants are all absorbed
into the coefficients $a_p$, where the sum can have a finite or an
infinite number of terms.  We only require that the integrals
converge.

The result for this generalised partition function was given in
\cite{JacLH} for degenerate masses and $N_1=N_2$. Our generalisation
follows \cite{AFV} closely. Because we differentiate the generalised
partition function with respect to the couplings $a_p$ in order to
generate expectation values, we have to keep track of all constants
that depend on the $a_p$.  The unitary matrix $U_0$ can be
parametrised as follows \cite{SV04,AFV}
\bea 
U_0 &=&  
\left( 
\begin{array}{cc} 
v_1 & 0\\ 
0   & v_2\\ 
\end{array} 
\right) 
\left( 
\begin{array}{cc} 
u_1 & 0\\ 
0   & u_2\\ 
\end{array} 
\right) 
\Lambda 
\left( 
\begin{array}{cc} 
v_1^\dagger &           0\\ 
0           & v_2^\dagger\\ 
\end{array} 
\right) \ , 
\nn\\
\Lambda&\equiv& 
\left( 
\begin{array}{ccc} 
\hat{\la}            & \sqrt{\one_{N_1}-\hat{\la}^2} & 0\\ 
\sqrt{\one_{N_1}-\hat{\la}^2} & -\hat{\la}            & 0\\ 
0                    & 0                    &-\one_{N_{2}-N_1}\\ 
\end{array} 
\right)\ . 
\label{Lapar} 
\eea 
Here, we denote by the matrix 
$\hat{\la}\equiv$ diag$(\la_1,\ldots,\la_{N_1})$ containing the real numbers 
$\la_k\in[0,1]$ for 
$k=1,\ldots,N_1$. The unitary submatrices are $u_1,v_1\in U(N_1)$, 
$u_2\in U(N_{2})$ 
and $v_2\in \tilde{U}(N_{2})\equiv U(N_{2})/(U(1)^{N_1}\times U(N_{2}-N_1))$.
The matrix $\Lambda$ is Hermitean and we observe that 
\be
\Tr\left[(U_0\Ga U_0^\dag \Ga )^p\right]\ =\ \Tr\Big[(\Lambda \Ga  \Lambda \Ga
  )^p\Big] 
\ ,
\label{udrop}
\ee
so that all unitary degrees of freedom drop out. From 
\be
(\Lambda \Ga  )^2= \left( 
\begin{array}{ccc} 
2\hla^2- \one_{N_1}            & -2\hla\sqrt{\one_{N_1}-\hla^2} & 0\\
2\hla\sqrt{\one_{N_1}-\hla^2} &  2\hla^2- \one_{N_1}                    & 0\\
0                              &  0                        & \one_{N_{2}-N_1}\\
\end{array} 
\right) ,
\ee
we obtain 
\be
\Tr\left[(U_0\Ga U_0^\dag \Ga )^p\right]\ =\ 
\Tr\left[
\left( 
\begin{array}{cc} 
2\hla^2-   \one_{N_1}          & -2\hla\sqrt{\one_{N_1}-\hla^2}\\
2\hla\sqrt{\one_{N_1}-\hla^2} &  2\hla^2-  \one_{N_1}          \\
\end{array} 
\right)^p\ 
\right]-\ \Tr[\one_{N_{2}-N_1}]\ .
\ee
The $2N_1$ eigenvalues 
can be written in diagonal matrix form as 
\be
X_\pm\ \equiv\ 2\hla^2- \one_{N_1} \pm i\,2\hla\sqrt{\one_{N_1}-\hla^2}  \ .
\ee
We thus arrive at 
\bea
\Tr\left[(U_0\Ga U_0^\dag \Ga )^p\right]+\ (N_{2}-N_1)\ 
&=& \Tr[X_+^p+X_-^p]
\ =\ \sum_{i=1}^{N_1} 2\sum_{q=0}^{[\frac{p}{2}]}
 {p \choose 2q}\Big(2\la^2_i-1\Big)^{p-2q}\Big(-4\la_i^2(1-\la_i^2)\Big)^q\ 
\nn\\
&=& \sum_{i=1}^{N_1}2T_{2p}(\la_i)\ .
\label{T2p}
\eea
The real polynomials $T_{2p}(\la_i)$ of degree $2p$ that we obtain are
the Chebyshev polynomials of the first kind\footnote{ In \cite{JacLH}
  the polynomial was given in the form $\cos(2p\cos^{-1}(\la))$.}.

Coming back to the integral in Eq.~(\ref{Zgen}), we can now go to an
eigenvalue basis using the parametrisation of Eq.~(\ref{Lapar}).
Because of the decoupling of
the unitary degrees of freedom in Eq.~(\ref{udrop}), the calculation is
identical to the one presented in \cite{AFV} Sec.~3, to which we
refer the reader for details.
In particular integrating out all unitary degrees of freedom cancels the 
Jacobian from the parametrisation eq. (\ref{Lapar}).
Collecting all of these results, we obtain the following
expression for $N_f$ flavours:
\be
{\cal Z}_{gen}(\{\eta\}) \ =\ 
\frac{\mbox{N}}{\Delta_{N_1}(\{\eta_{f1}\})\Delta_{N_2}(\{\eta_{f2}\})
} 
\left|
\begin{array}{ccc}
{\cal Z}_{gen}(\eta_{f1=1},\eta_{f2=1})&\cdots&
{\cal Z}_{gen}(\eta_{f1=1},\eta_{N_2})\\
\cdots & \cdots & \cdots\\
{\cal Z}_{gen}(\eta_{N_1},\eta_{f2=1})&\cdots&
{\cal Z}_{gen}(\eta_{N_1},\eta_{N_2})\\
I_\nu(\eta_{f2=1})&\cdots&I_\nu(\eta_{N_2})
\\
\eta_{f2=1}^{N_{2}-N_1-1}I_\nu^{(N_{2}-N_1-1)}(\eta_{f2=1})&\cdots&
\eta_{N_2}^{N_{2}-N_1-1}I_\nu^{(N_{2}-N_1-1)}(\eta_{N_2})\\
\end{array}
\right|,
\label{Zgenresult}
\ee
where $\Delta_{N_1}(\{\eta_{f1}\}=\prod_{j>i}^{N_1}(\eta^2_j-\eta_i^2)$
is the Vandermonde determinant within each flavour. 
The generalised $N_f=2$ flavour partition function, 
${\cal Z}_{gen}(\eta_1,\eta_2)$,
that is used inside the determinant above is given by 
\be
{\cal Z}_{gen}(\eta_1,\eta_2) \ \equiv\ 
\int_0^1 d\la\,\la\, \exp\left[ \sum_{p}a_p\Big( 2T_{2p}(\la)-(N_2-N_1)\Big) 
\right]
I_\nu(\la \eta_1)I_\nu(\la \eta_2)\ .
\label{Zgen2}
\ee
With $I_\nu^{(k)}(\eta_{f1})$ we denote the $k$-th derivative of the
$I$-Bessel function. The $I$-Bessel function itself, in fact, is a
one-flavour partition function, ${\cal Z}_\nu=I_\nu(\eta)$, 
up to a trivial $a_p$ dependent constant. 
The constant N in
Eq.~(\ref{Zgenresult}) is an irrelevant normalisation 
factor that does not contribute in
expectation values.
Eq.~(\ref{Zgenresult}) is the main new result of this appendix. It is the
solution of the generalised partition function given by the group
integral (\ref{Zgen}). It extends the previous result of \cite{JacLH}
to nondegenerate masses and $\Tr[\Gamma]\neq0$.

As an example, for $p=1$, we get 
\be
2T_2(\la)\ =\ 4\la^2-2\ ,
\ee
which gives back the standard partition function of Eq.~(\ref{ZNf2}), for 
$a_1=\frac12\al^2_{ef\!f}$ (and mass $\eta_{ef\!f}$). It is 
explicitly given in Eq.~(\ref{ZNf2}), up to a constant prefactor. For 
$p=2$, the second polynomial is
\be
2T_4(\la)\ =\ 16(\la^4-\la^2)+2\ ,
\ee
leading to the generalised partition function with $a_2=\omega$ that is 
needed in Eq.~(\ref{ZNf2beta}).


\end{document}